\documentclass[twocolumn,11pt, superscriptaddress, longbibliography]{revtex4-2}
\usepackage{graphicx}
\usepackage{amssymb}
\usepackage{amsmath,amsfonts,amssymb}
\usepackage{braket, MnSymbol}
\usepackage{bm}
\usepackage{commath}
\usepackage{cleveref}

\newcommand{\be}{\begin{equation}}
\newcommand{\ee}{\end{equation}}
\newcommand{\bea}{\begin{eqnarray}}
\newcommand{\eea}{\end{eqnarray}}
\newcommand{\ba}{\begin{array}}
	\newcommand{\ea}{\end{array}}

\newcommand{\bl}{\begin{flalign}}
\newcommand{\enl}{\end{flalign}}

\renewcommand{\bf}[1]{{\mathbf{#1}}}

\newcommand{\tdse}{time dependent Schr\"{o}dinger equation\ }

\newcommand{\eq}[1]{Eq. \eqref{#1}}

\newcommand{\intf}{\int_{-\infty}^\infty}
\newcommand{\mc}[1]{\mathcal{#1}}

\newcommand{\kket}[1]{|#1\rrangle}
\newcommand{\bbra}[1]{\llangle #1 | }
\newcommand{\bbraket}[1]{\llangle #1 \rrangle }
\begin{document}

\title{Sum rules for  light-dressed matter}
\author{Bing Gu}
\email{gubing@westlake.edu.cn}
\affiliation{
Department of Chemistry, School of Science, Westlake University, 600 Dunyu Road, Hangzhou 310030, Zhejiang Province, China}
\affiliation{ Institute of Natural Sciences, Westlake Institute for Advanced Study, 18 Shilongshan Road, Hangzhou 310024, Zhejiang Province, China}

\date{\today}

\begin{abstract}
	Light-driven matter can exhibit qualitatively distinct electronic and optical properties from those observed at equilibrium. We introduce  generalized sum rules for the optical properties of  driven systems by both quantum and classical light. For classical light, it shows that the  sum of all Fourier components, indexed by $n$, of the time-dependent dipole matrix elements between dressed states weighted by the corresponding quasienergy difference in the first Floquet Brillouin zone plus $n$ driving frequency is a constant, determined by the number of electrons. An analogous sum rule for quantum light-dressing is also derived.  These developments  provide guidance for the control of effective optical properties of matter by light fields.
\end{abstract}

\maketitle

\section{Introduction}

Matter ranging from atoms, through molecules to solid state materials, driven out of equilibrium by external stimulus, can have emergent electronic and optical properties that are qualitatively different from their equilibrium counterparts \cite{giovannini2019, oka2019, grifoni1998, bao2021, chu2004}. For example, laser-dressed atoms exhibit Autler-Townes effects and Mollow triplets \cite{Mollow1969}, dressed diatomic molecules  contain light-induced conical intersections \cite{natan2016}.  There are  a wealth of light-induced  phenomena for solid state materials \cite{seetharam2015, zenesini2009}  such as Thouless pumping \cite{Thouless1983}, modulated  optical susceptibility \cite{gu2018, Ghimire2011}, and tunable  band topology \cite{wang2013, schwennicke2020, wintersperger2020}. 

Optical driving provide a means to tune the optical properties of matter in an ultrafast femtosecond timescale.  
A particular interest is to employ far off-resonant drive, that can alleviate the heating induced by resonant pulses \cite{gu2018}. 
It has been experimentally demonstrated that ZnO driven by an infrared light shows below gap absorption  \cite{Ghimire2011}.
 A giant modulation of the second-order nonlinearity has been observed  for a layered magnetic insulator, manganese phosphorus trisulfide (MnPS$_3$) under  strong far off-resonant driving \cite{shan2021}. 

Understanding the emergent optical properties of driven matter poses challenges to the current theories that are developed under equilibrium conditions.
A general endeavor is to extend the concepts and theories that underlie our standing of equilibrium matter to non-equilibrium driven systems. Particularly, optical sum rules such as Thomas–Reiche–Kuhn (TRK) and Bethe sum rules \cite{bethe1930} play significant roles in our understanding of the optical properties of matter owning to its generality to all kinds of systems including atoms, molecules, and strongly interacting systems \cite{wang1999, chernyak1995}. The TRK sum rule is also important in the study of superradiant phase transition \cite{rzazewski1975} However, no such general rules have been found for dressed materials.   

Here we generalize the TRK sum rule to  systems   driven by external light. Both classical and quantum light are considered. The sum rule for classical light applies to any electronic material under periodic drive, and thus can provide useful insights into designing driven materials with desired effective optical properties. The sum rule for quantum light resembles closer the original TRK sum rule with the joint light-matter states replacing the electronic states.  These sum rules provide guidance to the manipulation of effective optical properties of electronic materials by light  fields, e.g., in Floquet engineering and cavity quantum electrodynamics \cite{wang2020a, garcia-vidal2021}. 

Atomic units $\hbar = e = m_\text{e} = 1$ are used throughout.

\section{Generalized TRK sum rule for Floquet matter}

 The general Hamiltonian for electronic materials interacting with a classical  light field is given by
$ H(t) = H_\text{M} + H_{\text{LM}}(t) $, where  $H_\text{M}$ is the electronic Hamiltonian
\be H_\text{M} = \sum_i \frac{\bf p_i^2}{2} + V(\bf r_i) + \sum_{i \ne j} V_{\text{ee}}(\bf r_i, \bf r_j)
,\ee
$\bf p_i$ is the momentum operator for $i$-th electron, 
$V(\bf r_i)$ the electron-nuclear interaction, and  $V_{\text{ee}}(\bf r_i, \bf r_j)$ the electron-electron Coulomb interaction , and $H_{\text{LM}}(t)  $ the light-matter interaction. In the electric-dipole gauge
\be H_{\text{LM}}(t) = -\bf d \cdot \bf E(t),
\ee
where $\bf d = -\sum_i \bf r_i$  is the dipole operator and $\bf E(t)$ is the total electric field of the dressing light. 
The driving can be strong to defy perturbative treatment of the light-matter interaction. 

%

A general  framework to understand strongly driven systems is provided by the Floquet theory.  
The Floquet theory provides not only an intuitive picture to understand driven matter but also an efficient computational method.   
For periodic Hamiltonian $H(t) = H(t+T)$, according to the Floquet theorem \cite{floquet1883, gu2018, Sambe1973}, there exits so-called Floquet states
\be \psi_\lambda(\bf r, t) = e^{-i\epsilon_\lambda t} \phi_\lambda(\bf r, t)
\label{eq:floquet}
\ee
that satisfy the \tdse. Here
$ \ket{\phi_\lambda(t)} = \ket{\phi_\lambda(t+T)} $ are the many-body Floquet modes, $T = 2\pi/\Omega$ is the driving period with $\Omega$ the driving frequency, and $\epsilon_{\lambda}$ are the quasienergies. The Floquet theorem is the analog of the Bloch  theorem in the time domain. 
It applies to a system driven by a laser with carrier frequency $\Omega$  together with any combination of its harmonics. 
Inserting \eq{eq:floquet} into the \tdse yields
\be \Gamma  \kket{\phi_\lambda}= \epsilon_\lambda \kket{\phi_\lambda}
\label{eq:gen_eig}
\ee
where $\Gamma = H(t) - i\pd{}{t}$ is the Floquet Hamiltonian.
Equation \eqref{eq:gen_eig} is a generalized eigenvalue equation defined in an extended Hilbert space,  a tensor product of the physical Hilbert space and the temporal space consisting of $T$-periodic functions.  In this so-called Sambe space, $\Gamma$ is  Hermitian and can be represented in a basis set $\set{\ket{\chi_\alpha} e^{i n \Omega t}}$ for numerical computations, where $\alpha$ runs over electronic states and $n = 0, \pm 1, \cdots$.  We use $\kket{\cdots}$ to denote states in the Sambe space. 
The electronic operator can be promoted to  the extended Hilbert space by 
$ A(\bf r, \bf p) \rightarrow A \otimes I $
where $I$ is the identity in time space.

An important relation between the  Floquet modes  is that if $\kket{\phi_\lambda}$ is a Floquet mode with quasienergy $\epsilon_\lambda$, then  $e^{-i n \Omega t } \phi_\lambda(\bf r, t)$ will also be an eigenstate of the Floquet Hamiltonian with quasienergy $\epsilon_\lambda + n \hbar \Omega$, i.e., $ $.  This can be  verified by inserting $e^{-i n \Omega t } \phi_\lambda(\bf r, t)$ into \eq{eq:gen_eig}.  Therefore, the full spectrum can be understood as the quasienergies uniquely defined in the first Floquet Brillouin zone (FFBZ) $[ -\hbar \Omega/2, \hbar \Omega/2)$ and their replicas with energies shifted by an integer number of driving frequencies.

To derive the sum rule, we start from the equality (assume the polarization in the $\hat{\bf z}$ direction)
\be \left[z_i, \left[ \Gamma, z_j \right]\right] = \delta_{ij} 
\label{eq:equality}
\ee
 which follows directly from the commutation relation $[ z_i,  p_z^{j}] = i \delta_{ij}$.
Multiplying both sides of \eq{eq:equality} by $e^2$ and sum over $i, j$ yields
\be   \left[d_z, \left[ \Gamma, d_z \right]\right] = N_\text{e} 
\label{eq:dipole}
\ee
where $d_z = \sum_i - e z_i $ is the dipole operator in $z$-direction, $N_\text{e}$ the number of electrons.


Inserting resolution of identity $\mc{I} = \sum_{\beta} \kket{\phi_\beta}\bbra{\phi_\beta}$ of the extended space in \eq{eq:dipole} and taking expectation value of state $\kket{\alpha}$ yields 
%

%

\be 2\sum_{\beta} \del{ \epsilon_{\beta} - \epsilon_\alpha } |\bbraket{\phi_\alpha | d_z |\phi_\beta}|^2 = N_\text{e} 
\label{eq:premain}
\ee
where $\epsilon_{\beta \alpha} = \epsilon_\beta - \epsilon_\alpha$.
Equation \eqref{eq:premain} is a sum rule that involves all  Floquet states in the Sambe space. However, 
 the number of Floquet states is infinity as $n \in (-\infty, \infty)$. 
It is more useful to focus on the Floquet states in the FFBZ.
To reduce the summation in \eq{eq:premain} to states only in the FFBZ, we use the relation 
\be \sum_{\beta} \epsilon_\beta = \sum_{\lambda \in \text{ FFBZ}} \sum_{n = -\infty}^{\infty} (\epsilon_\lambda + n \hbar \Omega ) 
\label{eq:111}
\ee
Let $\alpha = \lambda' $  be another Floquet mode in the FFBZ, the transition dipole matrix elements reads
\be   \bbraket{\phi_{\lambda'} | d_z |\phi_\lambda} =  \sum_n d^{(n)}_{\lambda'\lambda} 
\label{eq:112}
\ee
where $d^{(n)}_{\lambda'\lambda} = T^{-1}\int_{0}^{T} dt \bbraket{\phi_{\lambda'}(t) | d_z | \phi_{\lambda}(t)} e^{-in\Omega t}  $ is the $n$-th Fourier component of the time-dependent transition dipole between the Floquet modes. Such expansion is allowed 
due to the periodicity of the Floquet modes.


Using \cref{eq:111,eq:112} into \eq{eq:premain} yields 

\be 
2\sum_{\lambda \in \text{FFBZ}} \sum_{n = -\infty}^{\infty} (\epsilon_{\lambda \lambda'} + n \hbar \Omega ) \left|d^{(n)}_{\lambda'\lambda} \right|^2 = N_\text{e}  
\label{eq:main}
\ee
It can also be written as
\be \intf \frac{d\omega}{2\pi}  \omega S(\omega) = N_\text{e} 
\ee
where $S(\omega) = 2\pi \sum_{n = -\infty}^\infty \delta(\omega - \omega_{\lambda \lambda'} - n\Omega) |d^{(n)}_{\lambda'\lambda}|^2
$.

Equation \eqref{eq:main} is the sum rule for periodically driven materials. It is valid for strongly correlated materials and for  strongly driven matter.  
It reduces to the TKS sum rule (i.e. $ \sum_{\beta} |d_{\alpha \beta}|^2 2\epsilon_{\beta \alpha} = N $) in the equilibrium (non-driven) limit. For instance, consider a high-frequency driving where all the pristine eigenstates are contained in the FFBZ. The dressed states  become the eigenstates of $H_\text{M}$ and the inter-FBZ transitions are forbidden as the transition dipole between eigenstates are constants such that $n = 0$ is the only non-vanishing Fourier component.    

The sum rule provides a guideline for the manipulation of the optical absorption and emission properties of matter by lasers.  An intuitive way to understand \eq{eq:main} is that any two Floquet modes in the FFBZ can be associated with a set of transitions indicated by an integer number $n$. While it is tempting to consider an transition with $n \ne 0$ as photon-assisted transitions, this is only meaningful under a high-frequency driving while all pristine eigenstates reside in the FFBZ. In a low frequency driving $\Omega \ll E_g$ where $E_g$ is the characteristic energy gap of the pristine material, even the bare transitions will be associated with a large $n$ as  all pristine states folds into  a narrow spectral range. 
We can envision that the bare absorption spectrum of a material be broadened due to the replica structure of the quasienergies.

We now consider the case where the materials is driven by quantum light. Quantum light refers to electromagnetic fields with manifest quantum mechanical properties including entanglement as in entangled photon pairs generated through parametric down-conversion.  Coherent laser light is considered to be classical. Quantum light cannot be described by the semiclassical theory but  requires a full quantum electrodynamical approach. The full Hamiltonian then reads 
\be
\mc{H} =   \mc{H}_0  - \bf d \cdot \hat{\bf E} 
\ee 
where $\hat{\bf E}$ is the electric field operator at the molecule and $\mc{H}_0$ is the noninteracting Hamiltonian of matter and quantized field. 
Proceeding along the same lines as above, a similar sum rule is obtained 
\be 
2\sum_{\beta} \epsilon_{\beta\alpha} |\bbraket{\Phi_\alpha | d_z |\Phi_\beta}|^2 = N_\text{e}
\label{eq:main2}
\ee   
where $\kket{\Phi_\alpha}$, eigenstate of $\mc{H}$ with eigenenergy $\varepsilon_\alpha$, is a quantum light-dressed state, e.g., polariton state under strong light-matter coupling. Here $\kket{\cdots}$ refers to the joint light-matter space and $\beta$ runs over all eigenstates of the joint light-matter Hamiltonian, i.e., $\mc{H}_0$. 
\eq{eq:main2} is analogous to the non-driven TRK sum rule with the hybrid light-matter states replacing bare electronic states. It does not have the Floquet quasienergy structure as its spectrum is bounded from below.

To summarize, we have derived  generalized TRK sum rules for driven matter, valid for strongly driven systems.  While we have assumed a classical nuclei in the Hamiltonian, 
it can be easily generalized to quantized nuclei by replacing the electronic states with coupled electron-nuclear states.   This is due to that the nuclear kinetic energy operator does not change the commutation relation of electronic operators \cref{eq:equality}. Moreover, there can be  variants of its form if a different gauge is used and generalizations if a general operator is considered \cite{wang1999}.

\bibliography{../control,../optics,../cavity}
\end{document}